\documentclass[runningheads]{llncs}
\usepackage[T1]{fontenc}
\usepackage{graphicx}
\usepackage{amsmath,cite,url}
\usepackage{hyperref}
\usepackage{graphicx}
\usepackage{color}
\usepackage{tabularx}
\usepackage{enumitem}
\usepackage{censor}
\usepackage{algorithm}
\usepackage{algpseudocode}
\usepackage{amssymb}
\usepackage{bm}

\usepackage{lineno}
\begin{document}

\title{Steer-by-prior Editing of Symbolic Music Loops}
%
%
\author{Nicolas Jonason\orcidID{0009-0003-8553-3542} \and
Luca Casini\orcidID{0000-0002-3468-6974} \and
Bob L. T. Sturm\orcidID{0000-0003-2549-6367}}
\authorrunning{Jonason et al.}
%
\institute{KTH Royal Institute of Technology, Stockholm, Sweden
\email{\{njona,casini,bobs\}@kth.se}}
\maketitle              

\begin{abstract}
With the goal of building a system capable of controllable symbolic music loop generation and editing, this paper explores a generalisation of Masked Language Modelling we call Superposed Language Modelling. Rather than input tokens being known or unknown, a Superposed Language Model takes priors over the sequence as input, enabling us to apply various constraints to the generation at inference time. After detailing our approach, we demonstrate our model across various editing tasks in the domain of multi-instrument MIDI loops. We end by highlighting some limitations of the approach and avenues for future work.
We provides examples from the SLM across multiple generation and editing tasks at \href{https://erl-j.github.io/slm-mml-demo/}{https://erl-j.github.io/slm-mml-demo/}.
\end{abstract}


\section{Introduction}

Our goal is to build a system capable of controllable symbolic music loop generation \cite{han_loop,johnson_loopergp_2023} as well as editing on sparse representations of music.
The dominant paradigm for symbolic music generation uses Causal Language Models (CLM) conditioned on control signals such as a natural language descriptions, or specific control tokens \cite{johnson_loopergp_2023,ens_mmm_2020}.
The downside of CLMs is that they only generate in a left-to-right manner, making them difficult to adapt to editing tasks.

Masked Language Modelling generalises Causal Language modelling to allow sampling tokens in any order \cite{liao_probabilistically_2020-1,casini_investigating_2024,musebert,zeng_musicbert_2021}.
MLMs are appealing for editing tasks as they are inherently constrainable with unary equality constraints (an unmasked token at element $i$ constitutes an equality constraint for that element).
This gives us the ability to only regenerate certain aspects of a music piece like certain pitches, onsets, velocities while keeping others fixed.

\begin{figure}[ht]
    \centering
    \includegraphics[width=0.99\columnwidth]{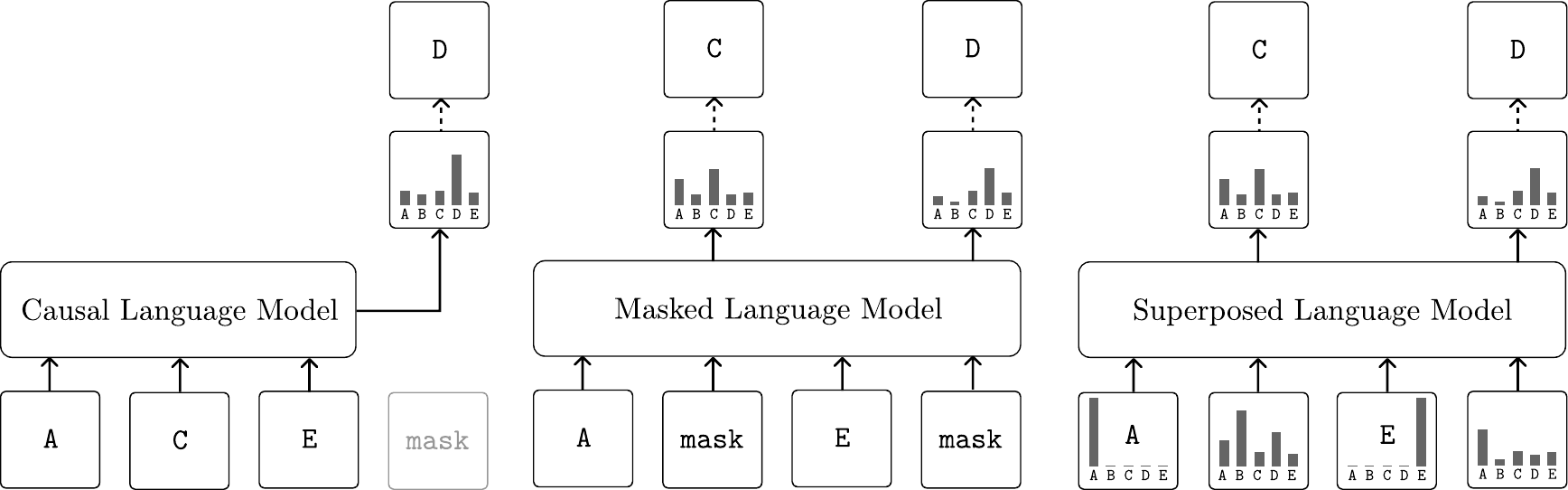}
    \caption{Causal language modelling, Masked Language Modelling and Superposed language Modelling of 4 letter words with a 5 letter vocabulary.}
    \label{fig:slm}
\end{figure}

We explore a further generalisation of Masked Language Modelling we call Superposed Language Modelling (SLM), seen in Figure \ref{fig:slm}. Rather than inputs being either known or unkown, SLMs instead receive a prior over the input sequence.
One advantage of this is that it allows us to impose constraints on a note attribute without fixing it to a particular value. 
For instance, we might want to restrict a xylophone's pitch to degrees of a particular scale, or have a hi-hat only on triplets. 

This paper presents experiments on training an SLM on a permutation invariant representation of 4-bar MIDI loops. We demonstrate the use of this model across various generation and editing tasks.\footnote{
    \href{https://erl-j.github.io/slm-mml-demo/}{\label{demo}https://erl-j.github.io/slm-mml-demo/}
}.

\section{MIDI Loop Data \& Representation}
We extract 4-bar multi-instrument MIDI loops from the MetaMIDI Dataset \cite{ens_metamidi_2021}, a dataset of multi-track MIDI files from a variety of styles. 
\paragraph{Splitting MetaMIDI Dataset}\label{sec:dataset}

We use $81\%$ for training, $9\%$ for validation and $10\%$ for testing.
To mitigate the chance of two or more MIDI files representing the same song ending up in different splits, we construct a bipartite graph with the MIDI file-hashes and the Spotify track IDs.
We then split the data according to connected components of this graph.
The MIDI files which have not been matched to Spotify tracks ids are assigned randomly to each split.

\paragraph{Extracting loops}
We adapt the book-ended-phrase loop detection heuristic from Adkins et al. \cite{johnson_loopergp_2023} to MIDI and extend it with additional metrical structure cues.
Under our extended heuristic a segment is considered a loop if the following is true:
i) 
it is book-ended by a musical phrase;
ii) its start coincides with an important metrical boundary;
and iii) its first beat coincides with the most important metrical boundary of the segment.
The reason for including metrical structure cues as an additional heuristic is that we find that the book-ended-phrase heuristic alone often yields loops whose first beat does not align with the perceived first-beat.
To detect the metrical structure, we use a pre-trained hierarchical metrical structure analysis model proposed by Jiang and Xia \cite{jiang_self-supervised_2023}. 
 
The final loop dataset consists of $\sim250,000$ 4-bar MIDI loops.

\paragraph{Representation}

The representation we choose has direct implications on the ease with which we can later implement different types of control.
Ordered representations of music bind information about the notes such as time, instrument and pitch to their absolute or relative position within the sequence.
With this in mind and inspired by \cite{musebert}, we use a \emph{permutation invariant} representation as it gives us the freedom to manipulate time related attributes of notes without consideration of their order. We represent a MIDI loop as a set of note event tuples. Each note event tuple contains $A=9$ attributes: instrument ($18$ instrument classes), pitch ($128$ for pitched instruments + $46$ for drums), onset beat ($0-16$), onset tick ($0-24$), offset beat ($0-16$), offset tick ($0-24$), velocity ($32$ levels), as well as the loop's global information tempo ($32$ levels) and style tags ($40$ tags). We represent \emph{inactive note events} by setting it's list of attributes to their respective undefined tokens.
We set the maximum number of notes of $N=300$, discarding longer loops and pad the remaining sequences with inactive notes.

\section{Implementing a Superposed Language Model}\label{sec:slm}

Our data consists of token sequences $x$ of length $T = NA$ with tokens from a vocabulary $V$. The SLM is written $Q = SLM(\pi)$. Its input is a matrix $\pi \in \mathbb{R}{[0,1]}^{T \times |V|}$ whose rows are categorical distributions over the token sequence representing some prior belief about the sequence $x$. Similarly, the output $Q \in \mathbb{R}{[0,1]}^{T \times |V|}$ is a matrix whose rows are categorical distributions representing the model's posterior belief about the sequence.

\paragraph{Superposition schemes}  Analogously to masking schemes in masked language modelling \cite{liao_probabilistically_2020-1}, we define a \textit{superposition scheme} $S$, i.e a procedure that takes an input sequence and generates a prior distribution, $\pi\sim S(x)$. For a superposition scheme to be valid, the following need to hold:
\begin{equation}
\label{sp-prop}
\forall \pi \sim S(x): \quad
0 \leq \pi_{t,v} \leq 1 \quad \forall t,v, \quad \sum_{v} \pi_{t,v} = 1 \quad \forall t, \quad \pi_{t,v} > 0 \text{ if } x_t=v
\end{equation}
where $\pi_{t,v}$ represents the probability of token $v$ at position $t$
and $x_t$ is the token at position $t$ in the input sequence.
In words, $\pi$ needs to consist of categorical distributions, i.e all probabilities are between $0$ and $1$ and sum to $1$.
Also, the ground truth sequence $x$ has to have non zero probability under the prior $\pi$. 

We experiment with a naive superposition scheme we call \texttt{Random-add} inspired by random masking schemes from masked language modelling \cite{liao_probabilistically_2020-1}.
\begin{algorithm}
\caption{\texttt{Random-add} superposition scheme}
\begin{algorithmic}[1]
\Require $X \in {0,1}^{T \times |V|}$ (one-hot representation of token sequence $x$)
\Require $\mathcal{S} \in {0,1}^{T \times |V|}$ (syntax mask)
\State Sample position masking probability $p_\text{pos} \sim \mathcal{U}(0,1)$
\State Generate binary position mask $M_\text{pos} \sim \text{Bernoulli}(p_\text{pos})^T$
\State Sample vocabulary masking probabilities $p_\text{vocab} \sim \mathcal{U}(0,1)^T$
\State Generate binary vocabulary mask $M_\text{vocab} \sim \text{Bernoulli}(p_\text{vocab})^{T \times |V|}$
\State Combine masks: $M \gets M_\text{pos} \odot M_\text{vocab}$
\State Apply OR operation: $X' \gets X \lor M$
\State Apply syntax mask: $\pi' \gets X' \odot \mathcal{S}$
\State Normalize: $\pi \gets \text{Normalize}(\pi')$
\Return $\pi$
\end{algorithmic}
\end{algorithm}
The general idea of this scheme is that it independently varies the ratio of unknown positions and the amount of noise we add to the unknown positions.
The reason for varying the ratio of unknown positions is to simulate the auto-regressive sampling used during inference, where the number of positions being known increases every decoding step \cite{liao_probabilistically_2020-1}.

The syntax mask $\mathcal{S}$ in step 7 is used to keep the priors syntactically valid. Since the note event tuples follow a fixed attribute order, we can ensure that the prior assigns zero probability for syntactically invalid sequences by multiplying the prior with a binary mask $\mathcal{S}$ where:
\begin{equation}
\mathcal{S}_{t,v} = \begin{cases}
1 & \text{if } v \text{ is syntactically valid at position } t \\ 0 & \text{otherwise}
\end{cases}
\end{equation}

\paragraph{Architecture}
We implement the SLM using a bi-directional transformer architecture \cite{devlin_bert_2019}.
We first embed the prior $\pi$ using a fully connected layer $W_{\text{emb}}$ without bias: $Z = W_{\text{emb}}\pi$ where $Z \in \mathbb{R}^{N A\times d}$.
This operation effectively computes a weighted sum of token embeddings, where the weights are the prior token probabilities at each position.
Following \cite{musebert}, we aggregate the embeddings within note events by summing each notes attribute embeddings, resulting in note event embeddings: $Z_{\text{note}}\in \mathbb{R}^{N\times d}$.
These note embeddings are then passed through a bi-directional transformer, allowing information exchange between note event embeddings:
\begin{equation}
Z_{\text{note}}' = \text{Transformer}(Z_{\text{note}})
\end{equation}
The resulting note output embeddings are then processed by a linear layer to produce note logits $l_\text{note}=W_{\text{out}}Z_{\text{note}}'$ where $l_\text{note} \in \mathbb{R}^{N \times |V|}$.
In order to get attribute specific logits, we repeat-interleave the note logits by a factor of $A$ to obtain $l\in \mathbb{R}^{NA \times |V|}$ and apply the syntax mask $\mathcal{S}$ with the following rule:

\begin{equation}
{l_Q}_{t,v} = 
\begin{cases}
-\infty & \text{if } \mathcal{S}_{t,v} = 0 \\
l_{t,v} & \text{otherwise}
\end{cases}
\end{equation}

\paragraph{Training \& hyperparameters}
We train the network using the cross-entropy loss of the output logits $l_Q$ w.r.t to the input sequence $x$.
We use a hidden size of 768, 12 layers, 12 attention heads and the norm-first configuration of the transformer. We train for 350k steps (4 days) on two \textit{NVIDIA GeForce RTX 3090}'s with a per-device batch size of $80$. We use the $Adam$ optimizer with a learning rate of $1e-4$ and epoch-wise learning rate decay of $0.99$.

\paragraph{Inference}
Starting from a prior, we sample unknown tokens autoregressively in random order, doing a full forward pass each time and applying a softmax to $l_{Q}$ to obtain token probabilities $Q$. For generating a loop, we need to do $T=NA$ forward passes in the worst case. We can optimize this with the following trick: set all of a note's attributes to "undefined" if any are sampled as such.

\begin{figure*}[t]
\includegraphics[width=\textwidth]{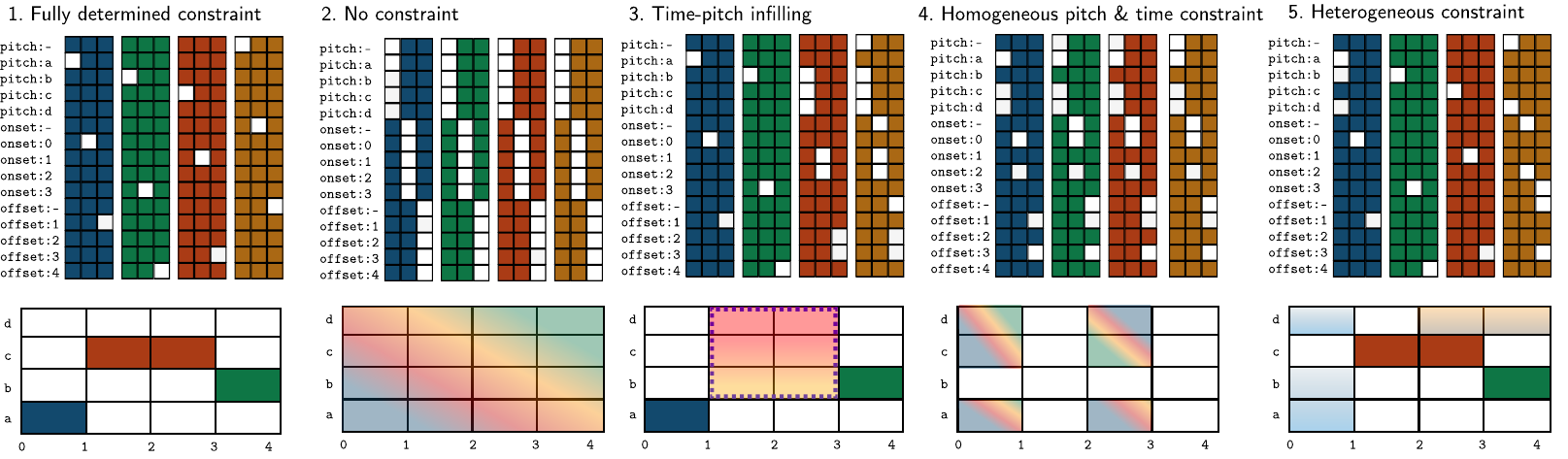}
    \caption{\textbf{Illustration of how priors on a unordered representations of music can be used to steer generation}. This example uses a toy-representation of music with 4 pitches, 4 onset times and 4 offset times. The upper row shows non-normalized vocabulary priors as binary masks where each column represents constraints on pitch, onset and offset respectively.
    The bottom row illustrates the constraints in piano roll form. 
    Each note event is colour coded. A colour gradient in the piano roll indicates that the note event(s) of the corresponding hue might be present in the region.
    \textbf{1.} shows a fully determined musical piece containing 3 notes. Notice how the orange note is inactive, hence all its attributes are set to undefined ("-").
    \textbf{2.} shows a completely unconstrained musical piece where nothing is known about any of the notes.
    \textbf{3.} shows a constraint representing a time-pitch infilling task. Notice how the red note and orange notes differ in their binary masks. Unlike red which is guaranteed to be active (all the "-" cells are 0), orange might or might not be active.
    This allows us to express precise ranges on the number of notes we want.
    \textbf{4.} shows how we can use constraints to express tonality and rhythm.
    \textbf{5.} shows how vocabulary constraints do not need to be uniform across the note events. Here, we only restrict the blue note's pitch and the orange note's onset.
    }
    \label{masks}
\end{figure*}

\paragraph{Steering generation with priors}
Figure \ref{masks} provides some toy examples of how priors can be used to steer generation. 

\section{Demonstration, Limitations \& Future work}

\paragraph{Demonstration} The URL \ref{demo} provides examples from the SLM across multiple generation and editing tasks. We also include examples from an MLM for comparison. We also provide a video showing an interactive application which uses the SLM to do various generation and editing tasks. Each action is executed by calling an API which first creates a synthetic prior based on the selected action, the current loop and predefined rules. The synthetic prior is then used as the input to the SLM to iteratively sample the unknown tokens.

\paragraph{Limitations} While we find that the steer-by-prior approach is a promising path for implementing various editing tasks, we also find some limitations.
First, the type of control we can exert is highly contingent on the chosen representation. For instance, our representation does not allow for direct control of the degree of polyphony.
Secondly, constructing the priors for each generation and editing task requires a combination of musical knowledge and programming knowledge.
Third, our SLM implementation is slow to sample despite it generating only 4-bars of music (it takes around 7 seconds to generate a loop on average).

\paragraph{Future work} Important avenues for future work are: 1) Rigorous objective and subjective evaluation of the system's outputs and usability and comparison with other approaches. 2) Improving sampling speed and extending the context-length without losing flexibility. 2) Improving usability by having an LLM provide a natural language interface to the system. 

%
%
\bibliographystyle{splncs04}
\bibliography{references}
\end{document}